\documentclass[reprint, superscriptaddress, nofootinbib]{revtex4-1}
\usepackage{color}
\usepackage[dvipsnames]{xcolor}
\usepackage{graphicx}

\usepackage{amsmath}
\usepackage{amsfonts}
\usepackage{amssymb}

\usepackage{multirow}
\usepackage{lineno}

\newcommand{\Hgen}{\ensuremath{\mathcal{H}}}
\newcommand{\Hpp}{\ensuremath{\mathcal{H}_{PP}}}
\newcommand{\Hnl}{\ensuremath{\mathcal{H}_{NL}}}

\newcommand{\Oppn}{\ensuremath{\mathcal{O}^{PP}_{N}}}
\newcommand{\Onln}{\ensuremath{\mathcal{O}^{NL}_{N}}}
\newcommand{\Onlpp}{\ensuremath{\mathcal{O}^{NL}_{PP}}}
\newcommand{\Otiger}{\ensuremath{\mathcal{O}^{!PP}_{PP}}}

\newcommand{\rhonet}{\ensuremath{\rho_\mathrm{net}}}
\newcommand{\rhorec}{\ensuremath{\rho_\mathrm{rec}}}

\newcommand{\f}{\ensuremath{f}}
\newcommand{\fref}{\ensuremath{f_\mathrm{ref}}}
\newcommand{\Hz}{\ensuremath{\mathrm{Hz}}}

\newcommand{\gmode}{$g$-mode}
\newcommand{\pmode}{$p$-mode}

\begin{document}


\title{
Impact of the tidal 
$p$-$g$ instability on the gravitational wave signal \\from coalescing binary neutron stars}

\author{Reed Essick}
\affiliation{Department of Physics and Kavli Institute for Astrophysics and Space Research, Massachusetts Institute of Technology, Cambridge, MA 02139, USA}
\affiliation{LIGO Laboratory, Massachusetts Institute of Technology, Cambridge, MA 02139, USA}
\author{Salvatore Vitale}
\affiliation{Department of Physics and Kavli Institute for Astrophysics and Space Research, Massachusetts Institute of Technology, Cambridge, MA 02139, USA}
\affiliation{LIGO Laboratory, Massachusetts Institute of Technology, Cambridge, MA 02139, USA}
\author{Nevin N. Weinberg}
\affiliation{Department of Physics and Kavli Institute for Astrophysics and Space Research, Massachusetts Institute of Technology, Cambridge, MA 02139, USA}

\begin{abstract}
Recent studies suggest that coalescing neutron stars are subject to a fluid instability involving the nonlinear coupling of the tide to $p$-modes and $g$-modes. 
Its influence on the inspiral dynamics and thus the gravitational wave signal is, however, uncertain because we do not know precisely how the instability saturates.  
Here we construct a simple, physically motivated model of the saturation that allows us to explore the instability's impact as a function of the model parameters.  
We find that for plausible assumptions about the saturation, current gravitational wave detectors might miss $> 70\%$ of events if only point particle waveforms are used.  
Parameters such as the chirp mass, component masses, and luminosity distance might also be significantly biased.  
On the other hand, we find that relatively simple modifications to the point particle waveform can alleviate these problems and enhance the science that emerges from the detection of binary neutron stars.
\end{abstract}

\maketitle


\section{introduction}\label{s:introduction}

The detection of gravitational waves (GWs) from binary black holes (BH)~\citep{GW150914,GW151226,O1BBH} with the Laser Interferometer Gravitational-wave Observatory (LIGO)~\citep{LIGO} opens a new window to our universe and provides the first tests of strong field general relativity (GR) in vacuum~\citep{GW150914testingGR,O1BBH}.
In the coming years, LIGO also expects to detect GWs from neutron stars (NSs) in coalescing binaries. 
Although a NS can be treated as a point particle (PP) to a first approximation, at some level tides will modify the rate of inspiral and thus the GW signal. 
The impact of the tidal effects are, however, uncertain.  
In part this is due to uncertainties in the NS equation of state, and indeed there is hope that GW observations will eventually provide precise constraints on the equation of state~\citep{Read:09, Hinderer2010, Damour:12, DelPozzo2013, Lackey2015, Agathos2015}.  
In addition, there are uncertainties in the tidal fluid dynamics both near the merger when matter and GR effects are strong~\citep{Read:13,Yagi:14, Favata:14} and during the long inspiral phase when the tide is weakly nonlinear \citep{Weinberg2013, Venumadhav2014, Weinberg2016}.

Many previous studies considered the impact of the linear tide, implicitly assuming that nonlinear effects are negligible at GW frequencies below $\f\approx 400\,\Hz$. 
These include studies of the linear equilibrium tide~\citep{Read:09, Hinderer2010, Damour:12, DelPozzo2013, Lackey2015, Agathos2015} and the linear dynamical tide in nonrotating NSs~\citep{Reisenegger:94,Lai1994, Hinderer:16, Steinhoff:16, Yu:16} and rotating NSs~\citep{Ho1999,Lai:06,Flanagan:07}. 
The equilibrium and dynamical tide refer, respectively, to the quasistatic and resonant response of a star to a tidal field (see, e.g., \citep{Ogilvie:14}).
Typically these studies conclude that linear tidal effects will be difficult to measure with current instruments without a gold-plated detection (signal-to-noise ratios $\gtrsim 50$; \cite{Read:09}) or stacked data from dozens of  marginal events~\citep{DelPozzo2013, Lackey2015, Agathos2015}.  
Moreover, because they find that tidal effects only become significant during the late inspiral, there are proposals to test vacuum GR using waveforms from NS systems at $\f\lesssim 400$ Hz~\citep{Agathos2014}.

Recently, it has been suggested that the tide is subject to a weakly nonlinear fluid instability during the early inspiral~\citep[][hereafter, VZH, W16, WAB,  respectively]{Weinberg2013,Venumadhav2014,Weinberg2016}.  
The instability involves a nonresonant coupling between the quasistatic equilibrium tide, pressure supported {\pmode}s, and buoyancy (i.e., gravity) supported {\gmode}s.  
Typically, modes first become unstable at $\f \approx 50\,\Hz$ and are driven thereafter to potentially large amplitudes. 
This continuous transfer of energy from the orbit into the modes increases the rate of inspiral and induces an evergrowing phase shift relative to the PP waveform. 
Although there has been disagreement in the literature about the magnitude of the growth rates, all studies of $p$-$g$ coupling predict an instability.
Furthermore, W16 find that nonstatic tidal effects (e.g., compressibility) enhance the growth rates, enabling a very large number of modes to reach significant amplitudes well before the binary merges.

Studies of the $p$-$g$ instability have mainly focused on calculating the instability threshold and growth rates;  they have not attempted to study its saturation in any detail. 
As a result, we do not know the rate at which the instability extracts energy from the orbit and thus we cannot say precisely how it will impact the GW signal.  
Because solving for the saturation is challenging and likely subject to uncertainties of its own, here we set a more modest goal.
We construct a parametrized model of the saturation and explore the instability's impact as a function of the model parameters. 
Our saturation model is relatively simple, adding just three new parameters to the 15 already present in the spinning PP model. 
It is worth emphasizing, however, that although we believe our saturation model adequately captures the range of possibilities, without a proper saturation study we cannot be certain.

The paper is structured as follows.
\S~\ref{s:physical mechanism} reviews the properties of the $p$-$g$ instability and discusses the physics of its saturation and the uncertainties therein.
\S~\ref{s:parameterized model} describes our parametrized model of the saturation which we use  to explore the tide-induced modifications to the PP waveform. 
Using Bayesian methods, which we describe in \S~\ref{s:bayesian analysis}, we then study how the modified waveforms affect source detectability and parameter bias if the tidal effects are neglected (\S~\ref{s:nlgr}) and how well we can measure the tidal effects if they are included (\S~\ref{s:nlnl}). 
We summarize and conclude in  \S~\ref{s:implications}.


\section{Nonlinear tidal instability}\label{s:physical mechanism}

As the NS inspirals and the amplitude of its tidal deformation increases, the tidal flow becomes susceptible to nonlinear fluid instabilities. 
These will initially manifest as weakly nonlinear interactions between the tide and internal oscillation modes of the star.  
WAB applied the formalism developed in~\citet{Weinberg:12} to determine the influence of such nonlinear interactions on the inspiral of NS binaries. This revealed a new form of nonlinear instability in which the tide excites a high-frequency \pmode~coupled to a low-frequency \gmode.  
Because the \pmode's (linear eigen-)frequency is much higher than the tidal frequency, the $p$-$g$ pair is not resonant with the tide.  
This form of three-wave interaction is therefore very different from the well-known resonant parametric instability in which the tide excites a pair of {\gmode}s whose frequencies approximately sum to the tidal frequency.\footnote{WAB showed that, although some {\gmode}s are also susceptible to the resonant parametric instability during the inspiral, their growth rates are too small to influence the GW signal.}

In analyzing the growth rates of the $p$-$g$ instability, WAB considered only three-wave interactions between the tide, a \pmode, and a \gmode. 
VZH showed that four-wave interactions between the tide (twice) and two {\gmode}s enter the analysis at the same order as the three-wave interactions. 
They found that the four-wave interactions significantly cancel the three-wave interactions and concluded that although the $m=\pm 2$ component of the equilibrium tide can be $p$-$g$ unstable, the growth rates are too small to influence the inspiral in a measurable way.

However, the analysis in VZH assumes that the equilibrium tide is incompressible.  
Although that is the case for the static equilibrium tide (the $m=0$ component), the nonstatic equilibrium tide ($m\pm2$) is compressible.  
W16 accounted for this compressibility and found that it undoes the cancellation between the three- and four-wave interactions, yielding rapid $p$-$g$ growth rates even during the early inspiral.  Specifically, W16 found that the instability turns on at gravitational wave frequencies \begin{equation}\label{eq:fgw_threshold}
    \f_{i} \simeq 45\left( \frac{\omega_g}{10^{-4}\lambda \omega_0} \right)^{1/2}\, \Hz,
\end{equation}
where $\omega_g$ is the \gmode's linear eigenfrequency, $\omega_0=(GM/R^3)^{1/2}$ is the dynamical frequency of a NS with radius $R$ and mass $M$, and $\lambda(a)\sim 0.1$--$1$ is a slowly undulating function of binary separation $a$ that depends on how close the (quasistatic) equilibrium tide is to a resonance (see Fig. 9 in W16).
On resonance, the tide is especially compressible and is more properly referred to as the dynamical tide. 

From Equation (\ref{eq:fgw_threshold}), we see that low frequency (i.e., high order) $g$-modes become unstable first.  
However, it is not clear what sets the minimum $\omega_g$ (the maximum $\omega_g$ is determined by the magnitude of the NS buoyancy frequency $\sim \omega_0/10$).  
W16 showed that, for $\omega_g \gtrsim 10^{-4} \omega_0$ (which corresponds to $\ell=2$ {\gmode}s with radial order $n\lesssim 10^{3}$), linear damping of the modes does not modify the instability threshold nor the growth rates. 
However, it is possible that other physical effects will limit the minimum $\omega_g$ (e.g., magnetic fields).  
As we describe in Section \ref{s:parameterized model}, our saturation model therefore includes a parameter that accounts for the uncertainty in $\f_i$.

\begin{figure*} 
    \includegraphics[width=1.0\textwidth]{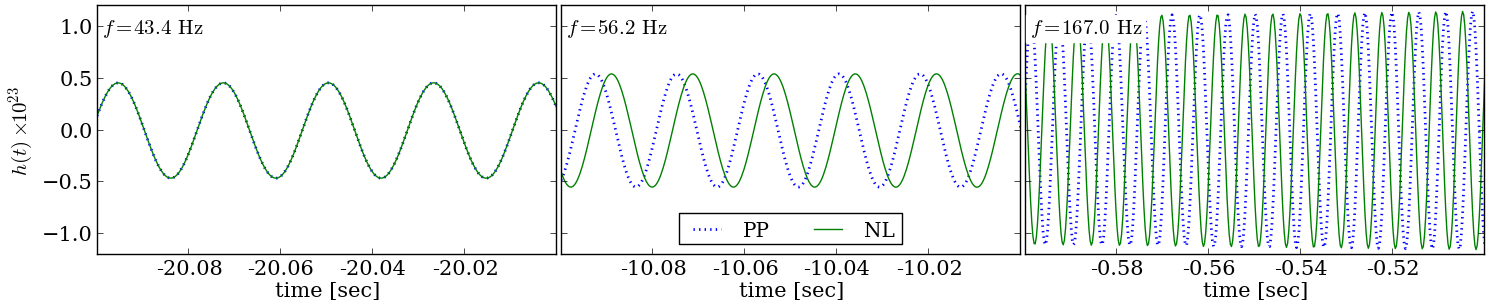}
    \caption{
        Time domain GW strain $h(t)$ for a 1.4$M_\odot$--1.4$M_\odot$ non-spinning binary NS system at three different stages of the inspiral.
        The blue dotted curves are the PP waveforms and the
        green solid curves are the waveforms with  nonlinear tidal effects assuming $A=4\times10^{-8}$, $f_0=50\,\mathrm{Hz}$, and $n=0$.
        }
    \label{f:h(t)}
\end{figure*}

Once unstable, the coupled {\pmode}s and {\gmode}s are continuously driven by the tide and their energy grows at a rate
\begin{equation}\label{e:Gamma_pg}
    \Gamma \approx 2\lambda \epsilon \omega_0 \simeq 20\lambda\left(\frac{M_2}{M_1+M_2}\right) \left( \frac{\f}{100\,\Hz} \right)^2\,\Hz,
\end{equation}
where $\epsilon=(M_2/M_1)(R_1/a)^3$ is the tidal amplitude parameter due to mass $M_2$ acting on mass $M_1$ and we assume $\omega_0=10^4\,\textrm{rad s}^{-1}$ [cf. Equation (112) in  W16; here we include an additional factor of 2 to yield the growth rate of the energy rather than the amplitude]. 
This equation is valid regardless of the relative size of the objects (i.e., both $M_1>M_2$ and $M_1<M_2$).
Note that  $\Gamma$ is independent of $\omega_g$, unlike $f_i$.  
Because the modes have enough time to grow by many tens of $e$-foldings before the binary merges (see W16 Section 5.4), eventually they reach such large energies that their growth saturates due to nonlinear damping (i.e., by exciting secondary waves through nonlinear wave-wave interactions).  
At saturation, there is a balance between continuous driving by the tide and decay through nonlinear damping.   
This suggests that the excited modes will continuously dissipate orbital energy at a rate
\begin{equation}\label{e:Edot_pg}
    \dot{E}_{NL} \approx \Gamma N E_{\rm sat}, 
\end{equation}
where $N$ is the number of independently unstable modes.  
The value of $N$ is uncertain, but because the modes do not need to be resonant, $N\sim 10^3$--$10^4$ is possible based on the modes' typical radial order and angular degree ($n\sim 1000$, $\ell \sim \textrm{ few}$). 

Given $\dot{E}_{NL}$, we can calculate the cumulative phase shift of the GW signal relative to the PP signal (see Appendix~\ref{s:derivation of phase shift} for details)
\begin{equation}\label{e:dphi_pg}
    \Delta \phi(\f) \approx  2\pi \int_{\f_i}^{\f} \frac{\dot{E}_{NL}}{\dot{E}_{\rm gw}} \tau d\f,
\end{equation}
where $\dot{E}_{\rm gw}$ is the GW luminosity, $\tau=\f/\dot{\f}$ is the inspiral time scale (both of which are dominated by the leading order quadrupole formula for two point masses~\citep{Peters1963}), and $\dot{\f}$ is the rate at which the gravitational-wave frequency increases with time. Note that if the binary contains two NSs, the instability manifests in each star separately and their individual $\dot{E}_{\rm NL}$ add to the system's total $\Delta \phi$.

In general, $E_{\rm sat}$ will be a complicated function of $\Gamma$, $N$, the properties of the unstable modes, the NS structure, and the equation of state. 
Calculating $E_{\rm sat}$ is therefore challenging and beyond the scope of this paper.  
Nonetheless, we might expect wave breaking to set an approximate upper bound. 
A wave breaks when $k_r\xi_r \sim 1$, where $\xi_r$ is the amplitude of the wave's radial displacement and $k_r$ is its radial wave number.  
At wave breaking, a \gmode~overturns the local stratification and a \pmode~induces order unity density perturbations.  
WAB show that {\gmode}s in a NS break at an energy
\begin{equation}\label{eq:Ebreak}
    E_{\rm break} \sim 10^{-8} \left( \frac{\omega_g}{10^{-4}\Lambda_g \omega_0} \right)^2 \left( \frac{r}{R} \right)^2 E_0,
\end{equation}
where $r$ is the radial location within the star at which the breaking occurs, $\Lambda_g = \ell_g (\ell_g+1)$,  and $E_0=GM^2/R$.   
This is lower than the energy at which the {\pmode}s break and thus the {\gmode}s probably determine $E_{\rm sat}$ for the $p$-$g$ instability. 
Although we use $E_\mathrm{break}\sim10^{-8}E_0$ as a reference value throughout our study, note that if  the {\gmode}s break at $r\ll R$ the actual value will be much smaller.

These considerations motivate the ansatz $E_{\rm sat} = \beta E_{\rm break}$, where $\beta\rightarrow1$ corresponds to saturation at the \gmode~wave breaking energy.  
Observations of \gmode~instabilities in the ocean, the atmosphere, and laboratory experiments often find that saturation indeed occurs by wave breaking (see the review by Staquet and Sommeria~\citet{Staquet:02}).  
Numerical studies of the dynamical tide in hot Jupiter systems find that {\gmode}s driven by the parametric instability also saturate at energies $E_{\rm sat} \sim E_{\rm break}$ \citep{Barker2010, Barker2011, Essick2016}.  
This suggests that perhaps $\beta\sim 1$ for the $p$-$g$ instability as well.  

To summarize, $\dot{E}_{NL}$ and therefore $\Delta \phi$ are poorly constrained because of uncertainties in the minimum $\omega_g$, the number of unstable modes $N$, and the saturation energy $E_{\rm sat}$ (or equivalently, $\beta$). 
We now describe how our parametrized model of the saturation accounts for these uncertainties.


%

\section{parametrized model of the saturation}\label{s:parameterized model}

While the saturation of the $p$-$g$ instability is likely to be a complicated process, we construct a relatively simple model motivated by the theoretical considerations discussed in \S~\ref{s:physical mechanism}.
Given Eqs. (\ref{e:Gamma_pg}) and (\ref{e:Edot_pg}), we  model the saturation with three parameters ($A$, $\f_0$, $n$) such that  
\begin{equation}
    \dot{E}_{\rm NL} \propto \lambda f^2 N E_{\rm sat} \propto A f^{n+2}\Theta\left(f-f_0\right),
\end{equation} 
where $\Theta$ is the Heaviside function.  
The model assumes that $\beta N\lambda \propto f^n$ for $f>f_0$.  
The parameters $A$ and $n$ determine the overall amplitude and frequency dependence of $\dot{E}_{\rm NL}$ while $f_0$  is the frequency at which the modes reach saturation.
By allowing $A$, $f_0$, and $n$ to vary, we can account for the  uncertainties in $f_i$, $\lambda$, $N$, and $E_{\rm sat}$ discussed in \S~\ref{s:physical mechanism}. 
 In Appendix~\ref{s:derivation of phase shift} we show that
\begin{eqnarray}
A &=& \left(\frac{2\pi \fref}{\omega_0}\right)^{1/3}\left(\frac{\omega_g}{\Lambda_g \omega_0}\right)^2 \left[\beta N \lambda\right]_{\rm ref}
\nonumber \\
&\simeq& 4\times10^{-9}\left(\frac{\omega_g}{10^{-4}\Lambda_g \omega_0}\right)^2\left[\beta N \lambda\right]_{\rm ref},
\label{eq:Aparam}
\end{eqnarray}
where $\fref$ is a reference frequency that sets the dimensionless scale of A but is otherwise arbitrary and $\left[\beta N \lambda\right]_{\rm ref}$ indicates the value of $\beta N\lambda$ at $\f=\fref$.
 We choose $\fref=100\textrm{ Hz}$ throughout our study.
  Note that our model ignores any dissipation that might occur when $f_i< f < f_0$  and instead assumes that $\dot{E}_{\rm NL}$ turns on as a step function at $f_0$ (such discontinuities can cause problems for Fisher-matrix studies~\citep{Mandel2014} but not for our analysis because we do not differentiate the GW phase).

By Eq. (\ref{e:dphi_pg}), the cumulative phase shift due to the tide raised in $M_1$ by $M_2$ is then (see Appendix~\ref{s:derivation of phase shift}), 
\begin{eqnarray}\label{e:phase shift}
    \Delta \phi (x>x_0)  & =     & A F_{M} \left[\frac{x_0^{n-3}-x^{n-3}}{n-3}\right] \nonumber \\
                         &\simeq & 0.4 \left(\frac{\mathcal{M}}{1.2 M_\odot}\right)^{-10/3} \left(\frac{A}{10^{-8}}\right)\left[\frac{x_0^{n-3}-x^{n-3}}{n-3}\right] \textrm{rad} \nonumber \\
\end{eqnarray}
where $x=\f/\fref$, $x_0=f_0/\fref$, $F_M$ is given by
\begin{equation}\label{e:Fm}
    F_M = \frac{25}{1536} \left(\frac{2M_1}{M_1+M_2}\right)^{2/3} \left(\frac{G\mathcal{M} \pi \fref}{c^3}\right)^{-10/3},
\end{equation}
and $\mathcal{M}$ is the chirp mass [$\mathcal{M}=(M_1M_2)^{3/5}/(M_1+M_2)^{1/5}$]. 
The numerical result on the second line assumes a NS-NS binary with $M_1=M_2$ and $n<3$ (note $\mathcal{M}\simeq 1.2 M_\odot$ for $M_1=M_2=1.4M_\odot$).
We again note that this expression is valid both when $M_1>M_2$ and when $M_1<M_2$.

The total phase shift accumulated by the time the NS merges is $\Delta \phi (f\gg f_0) \propto A  f_0^{n-3}$. 
Because the growth rates are large compared to the inspiral time, $f_0 \simeq f_i$ and thus $\Delta \phi (f\gg f_0) \propto \omega_g^{(n+1)/2}$, assuming $[\beta N \lambda]_{\rm ref}$ is independent of $\omega_g$. 
Because we expect $n>-1$, we see that unstable modes with \emph{larger} $\omega_g$ contribute more to $\Delta \phi$ at merger (as long as $\omega_g$ is sufficiently small that the modes reach saturation before the merger).  
This is because modes with smaller $\omega_g$  have smaller $E_{\rm break}$ [Eq. \ref{eq:Ebreak}] and thus contribute less to the total $\dot{E}_{\rm NL}$ despite being unstable earlier in the inspiral [Eq. \ref{eq:fgw_threshold}].
 
The phase shift depends on the component masses as $\Delta \phi \propto  [1+q]^{-2/3}\mathcal{M}^{-10/3}$, where $q=M_2/M_1$ is the mass ratio.\footnote{Normally, we only consider $q\leq1$ because of a symmetry under the interchange $M_1\leftrightarrow M_2$, but this is not the case for $\Delta\phi$ caused by only the tide in $M_1$ raised by $M_2$. If we included the phase shift induced by both the tide in $M_1$ raised by $M_2$ and vice versa, as we do later in this study, the symmetry is restored.}
Highly asymmetric systems, such as NS-BH binaries, therefore have much smaller $\Delta \phi$ all else being equal.  
This is because NS-BH orbits decay faster and there is less time for the nonlinear tidal effects to accumulate during the early inspiral.
For example, $\Delta \phi$ is approximately $100$ times smaller for a NS-BH binary with a $1.4 M_\odot$ NS and a $10M_\odot$ BH compared to an NS-NS binary with $M_1=M_2=1.4 M_\odot$ (accounting for the $\Delta \phi$ due to both NSs). 
As we describe below, we expect $A\lesssim 10^{-6}$ and a NS-BH binary has $\Delta \phi \lesssim 1 \textrm{ rad}$.
We show in \S~\ref{s:detectability} that such a phase shift is at the margins of detectability.

In our analysis, we consider values of $A$  in the range $10^{-9} \lesssim A \lesssim 10^{-6}$.  
From Eq. (\ref{eq:Aparam}) we see that $A\sim 10^{-6}$ corresponds to, e.g., $N\sim 10$ ($\sim 10^3$) modes with  $\omega_g /\omega_0\sim 10^{-3}$ ($\sim 10^{-4}$) each saturating near their wave breaking energy $\beta \sim 0.1$--$1$.  
These values of $N$ are based on the radial and angular orders of such modes ($n\sim100$--$1000$ and $\ell\approx \textrm{ few}$).  
We therefore do not expect $A$ to be much larger than $10^{-6}$. 
Regarding the low end of our $A$ range, we will show that for $A\lesssim10^{-8}$ the phase shift is too small to be detectable. 

Because we do not expect $\dot{E}_{\rm NL}$ to be a particularly strong function of $f$, we consider values for $n$ in the range $0 \le n \le 2$. 
As the binary separation decreases, higher frequency modes become unstable  [Eq. \ref{eq:fgw_threshold}], which suggests that $N$ and perhaps $E_{\rm sat}$ increase with $\f$, implying $n>0$. 
Finally, the rapid growth rates suggest that $f_0$ is close to $f_i$. 
We therefore consider values in the range $30 \lesssim f_0 \lesssim 80\,\Hz$.
     
\begin{figure} 
    \includegraphics[width=1.0\columnwidth]{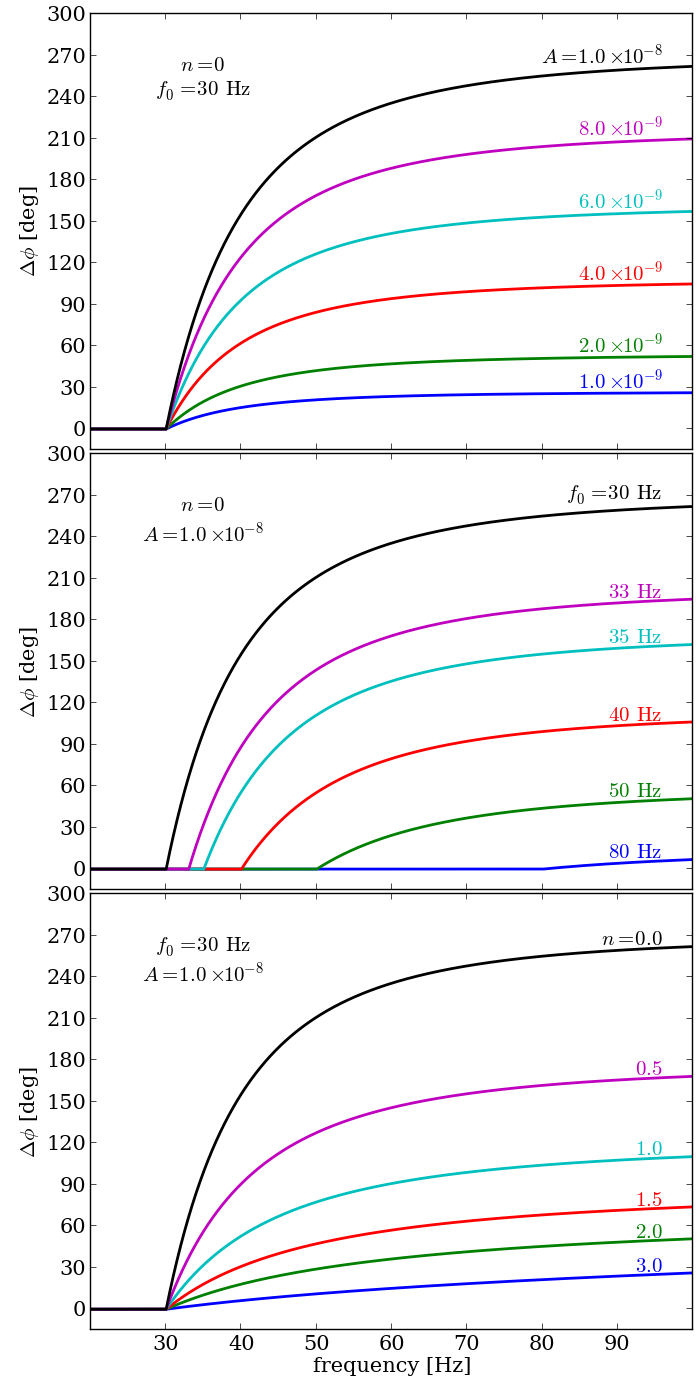}
    \caption{
        Cumulative phase shift $\Delta \phi$ as a function of GW frequency $f$ and its dependence on the model parameters  $A$, $f_0$, and $n$ (top, middle, and bottom panels, respectively).
        }
    \label{f:parameterized model}
\end{figure}

The saturation parameters may depend on the stellar structure and thus the component masses.  
We therefore allow each star in a binary NS system to have its own $A$, $n$, and $\f_0$.
Following previous work~\cite{DelPozzo2013}, we expand all the parameters around a reference mass. 
To wit, 
\begin{equation}\label{e:taylor}
    A(M_i) \equiv A^{(0)} + A^{(1)} (M_i -1.4 M_\odot) + \cdots.
\end{equation}
In our analyses, we keep only the zeroth and first order terms and although we marginalize over both orders, we focus on the zeroth order terms throughout this study, dropping the superscript unless otherwise indicated. 
For simplicity we consider only the mass dependence; future improvements to the model might allow for dependencies on other stellar parameters (e.g., spin and magnetic fields).

In Fig.~\ref{f:h(t)} we show the time domain waveform, with and without the nonlinear corrections to the PP solution, assuming a nonspinning $1.4 M_\odot$--$1.4 M_\odot$ binary NS system. 
Although both waveforms are identical at early times ($f < f_0$), a phase shift accumulates throughout the inspiral.
For these parameters,  the cumulative phase shift at $f\simeq 200\Hz$ is $\Delta \phi \simeq 200^\circ$. As we show in \S~\ref{s:detectability},  the nonlinear tidal effects begin to be detectable at such phase shifts.  

In Fig.~\ref{f:parameterized model} we show $\Delta \phi(f)$ for a range of $A$, $\f_0$, and $n$.  
Large $A$ implies large total phase shift $\Delta \phi(f\rightarrow\infty)$ whereas large $f_0$ or $n$ imply the opposite. 
We also see that although $\Delta \phi(f\rightarrow\infty)$ depends on all three parameters, the slope is mostly determined by $A$ and $n$.  
Moreover, because we expect $n<3$, $\Delta \phi$ accumulates most rapidly at low frequencies and asymptotes to a constant value at large frequencies.  Since the PP models can account for a constant overall phase shift, detecting the nonlinear tidal effects depends primarily on the low-frequency sensitivity of the detectors.

Assuming a parameterized post-Einsteinian formalism, ~\citet{Cornish2011} study  modifications to PP GR waveforms that are, in some ways, similar to ours.  
In particular, they assume a power-law form for the phase shift, $\Delta\phi(f) \sim Af^n$, and explore a range of power-law amplitudes and exponents. 
However, they do not include a turn-on frequency $f_0$.  
Furthermore, they focus on high frequencies because they find that solar-system tests are more sensitive to deviations from GR than GW measurements at low frequencies. Nonetheless, their conclusions are consistent with ours to the extent that they can be compared. 


\section{Bayesian inference}\label{s:bayesian analysis}

We use Bayesian methods to assess how our model of the nonlinear tidal effects impacts the GW data analysis.  Specifically, we use \textsc{Nested Sampling}~\citep{Skilling2006,Veitch2010} within \textsc{LALInference}~\citep{LALInference} to compute posterior distributions and the evidence.
In the most general PP case, the GW signal emitted by a binary in a circular orbit depends on 15 parameters, including the two component masses, source location, orientation, distance, and 6 degrees of freedom for the two spins.
We collectively refer to the unknown parameters as $\vec{\theta}$. 
In a Bayesian framework, the evidence $Z$ of data $d$ given a model \Hgen{} is
\begin{equation}\label{e:Evidence}
    Z\equiv p(d|\mathcal{H}) = \int d\vec{\theta}\, p(d| \vec{\theta},\mathcal{H}) p(\vec{\theta}|\mathcal{H}),
\end{equation}
where the first term in the integral is the likelihood and the second is the prior, both of which depend on the model.
The multidimensional posterior distribution of $\vec{\theta}$ can be written using Bayes' theorem as
\begin{equation}\label{e:Posterior}
    p(\vec{\theta} | d, \mathcal{H})= \frac{ p(d| \vec{\theta},\mathcal{H}) p(\vec{\theta}|\mathcal{H}) }{Z}.
\end{equation}
Furthermore, if two (or more) competing models are available, odds ratios between pairs of models can be calculated as
\begin{equation}
    \mathcal{O}^A_B = \frac{p(\mathcal{H}_A|d)}{p(\mathcal{H}_B| d)} = \frac{p(\mathcal{H}_A) Z_A}{p(\mathcal{H}_B) Z_B},
\end{equation}
where the ratio of priors reflects the initial relative belief in each model. 
We assume that  no model is preferred a priori and therefore $\mathcal{O}^A_B\rightarrow Z_A/Z_B$.

When the gravitational waveform's shape is known a priori, we use templates to represent the expected signal.
These templates are parameterized by $\vec{\theta}$ and form a manifold onto which we project the data.
By measuring how well different points on the manifold match the data, we construct posterior distributions for each signal parameter.
This is effectively what is done within Eqs.~(\ref{e:Evidence}) and~(\ref{e:Posterior}).
However, if the manifold does not accurately capture the full range of possible signals, biases may be introduced. 
Furthermore, if no point on the manifold represents the data well, we may not be able to recover the signal at all (small $Z$).
This effect, commonly referred to as template mismatch, can occur if the phase shift introduced by nonlinear tides is sufficiently large and neglected.

In what follows, we consider two models: \Hpp~treats the two objects as point particles, whereas \Hnl~includes nonlinear tidal effects.
The \Hpp~model uses a simple inspiral-only analytic approximant (TaylorF2)~\citep{WaveformComparison}. 
The \Hnl~model augments the TaylorF2 phase evolution with a tide-induced phase evolution given by Eq.~(\ref{e:phase shift}).

We focus on a single, optimally oriented, nonspinning\footnote{We briefly consider aligned spins in \S~\ref{s:detectability}.} binary NS system, analyzed at distances corresponding to network signal-to-noise ratios \rhonet~near 12, 25 and 50. 
These roughly correspond to marginal, confident, and gold-plated detections, respectively.
We also neglect linear tides, which we expect to decouple from the NL effects because the former are significant at high frequencies while the latter are most significant at relatively low frequencies (see Fig. \ref{f:parameterized model}).
We include the LIGO Hanford and Livingston detectors in addition to Virgo~\citep{Virgo}, assuming expected sensitivities for the second observing run (O2)~\citep{ObservingScenarios}. 
While these may not be realized exactly, they should approximate the relative sensitivities of the detectors. 
Because detections will be driven by the two LIGO instruments, which are expected to be more sensitive than Virgo, we place our signal directly overhead North America~\citep{ObservingScenarios}.
Virgo will mostly just improve localization through triangulation,  although it could also help constrain intrinsic parameters for loud, precessing systems through improved polarization constraints.
Finally, we use a zero-noise realization for our simulations, which is equivalent to taking the expected value of the evidence and posterior distributions from many noise realizations~\citep{Vallisneri:08}.
Details of our priors on all parameters are provided in Appendix~\ref{s:priors}.


\section{detectability and biases when nonlinear tides are neglected}\label{s:nlgr}

We begin by investigating the impact of neglecting nonlinear tidal effects.
We do this by injecting signals that include the tide-induced phase shift but then fit the data using only the PP waveforms.
This causes significant template mismatch if the tidal effects are large,  impairing our ability to detect events and biasing the inferred parameters.

Detectability and bias are related but subtly different~\citep{Lindblom2008,Cutler2007}. 
For example, the best fit may not be very good but nevertheless remain near the true parameters (i.e., unbiased but impaired detection). 
Alternatively,  we may be able to find a good fit  but only with parameters that are far from the true values (i.e., biased but unimpaired detection).  
Depending on the magnitude of $\Delta \phi$ and its frequency evolution, we observe one or both effects.  


\subsection{Detectability}\label{s:detectability}

As $\Delta \phi$ increases, the template mismatch worsens. 
We generally find that when $A\gtrsim10^{-8}$ nonlinear tidal effects begin to be noticeable for current detector sensitivities. 
From Fig. \ref{f:parameterized model}, we see that this corresponds to $\Delta \phi \gtrsim 1\textrm{ rad}$, which is similar to other estimates of the minimum measurable $\Delta \phi$ (e.g., \cite{Cutler:94, Balachandran:07}).
In terms of the saturation model described in \S~\ref{s:parameterized model} [see Eq. (\ref{eq:Aparam})], $A\sim10^{-8}$ corresponds to, e.g., $N\sim 10$ unstable modes with $\omega_g\sim 10^{-4} \omega_0$ saturating at $E_{\rm sat}\sim E_{\rm break}$ or equivalently $N\sim 10^3$ such modes saturating at $E_{\rm sat} \sim 0.01 E_{\rm break}$.

\begin{figure}
    \includegraphics[width=1.0\columnwidth]{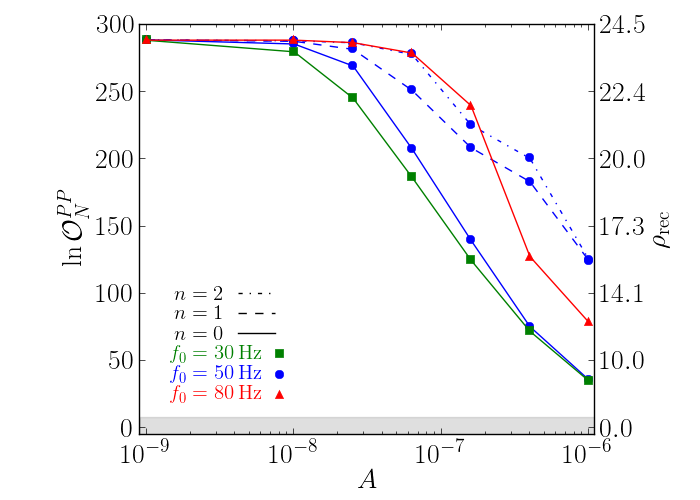}
    \caption{
       Odds ratio \Oppn~for injected signals that include nonlinear tidal effects but are recovered using PP waveforms.
        The signals are injected at $\rhonet\simeq25$.    
        The right axis shows the recovered signal-to-noise ratio \rhorec, computed from \Oppn~in the Laplace approximation as $\Oppn = \rhorec^2 / 2$.
        }
    \label{f:nlgr logBSN}
\end{figure}

We illustrate this result in Fig.~\ref{f:nlgr logBSN} for signals that include nonlinear tidal effects injected with $\rhonet\simeq25$.  
We show the odds ratio \Oppn~of a PP waveform model relative to pure Gaussian noise as a function of $A$ for different values of $n$ and $f_0$. 
For small $A$, \Oppn~plateaus at large values because the PP signal model matches the data well.
However, as $A$ increases the PP model matches the data less and less, thereby decreasing the evidence for the existence of a signal.
\Oppn~can be mapped into the recovered \rhonet~(called \rhorec), and we see that for  $A\sim 10^{-6}$ more than half of the signal is lost ($\rhorec < \rhonet/2$). 
In that case, the horizon distance shrinks in half and we miss approximately $1-(1/2)^3\simeq 90\%$ of NS merger events. 
For $\rhonet\simeq12$, extreme values of $A$  can produce $\Oppn<1$, which implies that Gaussian noise is preferred over the PP signal model even though we use a zero-noise realization.

We injected similar signals with three different \rhonet~(12, 25, and 50), although we only show the results for $\rhonet\simeq25$ because we find that all \rhonet~yield very similar results modulo the usual broadening of posteriors associated with lower \rhonet~signals.
  For example, all \rhonet~produce nearly identically shaped \Oppn~curves and simply scale \Oppn~up or down.
Signal loss due to template mismatch produces this behavior because we lose a fixed fraction of the inner product between the template and the data regardless of the overall amplitude.

\begin{figure}
    \includegraphics[width=1.0\columnwidth]{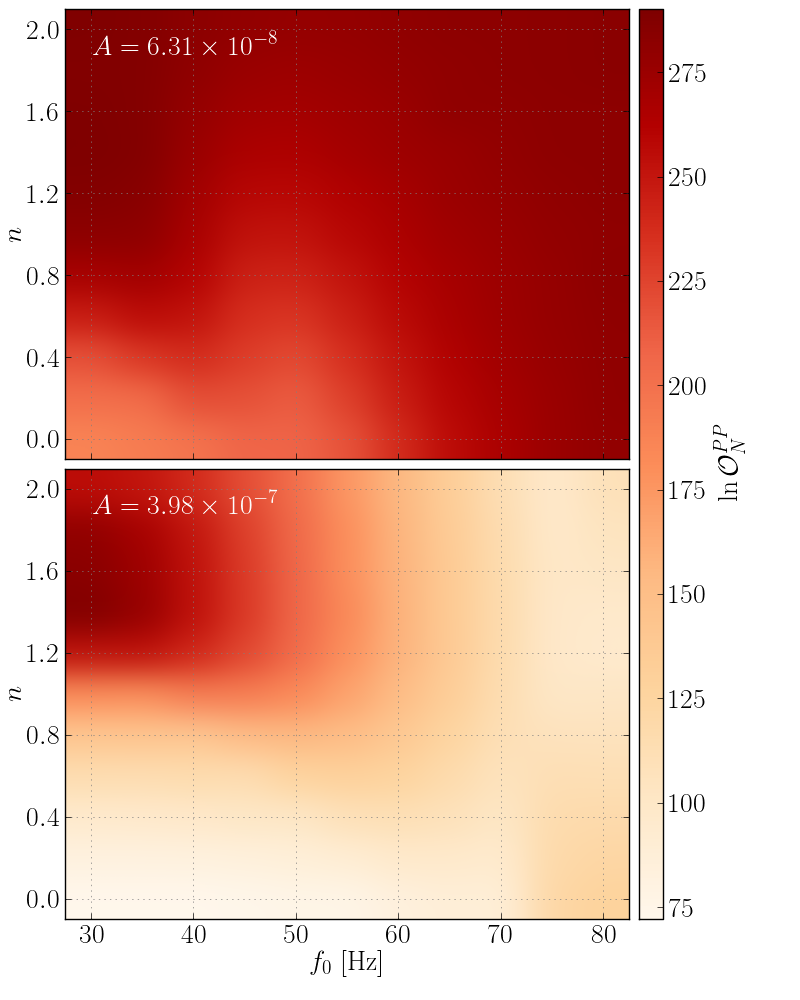}
    \caption{
         Surface plots of  \Oppn~as a function of $n$ and $f_0$ at  $A=6.3\times10^{-8}$ (top panel) and  $A=4.0\times10^{-7}$ (bottom panel).
         The signals are injected at $\rhonet\simeq25$.
        }
    \label{f:nlgr logBSN surf}
\end{figure}

As Fig.~\ref{f:nlgr logBSN} shows, the decrease of \Oppn~ with increasing $A$ depends only mildly on $n$ and $f_0$. 
We can see this in more detail in Fig.~\ref{f:nlgr logBSN surf}, which shows \Oppn~as a function of $n$ and $f_0$ for two values of $A$. 
Typically, small $n$ imply more rapid accumulation of phase shift and small $f_0$ imply more total phase shift, both of which produce larger template mismatch and lower \Oppn. 
We also see that \Oppn~depends more strongly on $A$ for high $\f_0$ injections than for low $\f_0$ injections.


\subsubsection{Effects of spin}\label{s:spins}

We also briefly investigated the effects of spins with TaylorF2 approximants.
These signals allow the components to spin either aligned or antialigned with the orbital angular momentum, and therefore do not include precession effects.
Spins can change the waveform's duration, which may be confused with the analogous effect from nonlinear tidal interactions.
Searches often use TaylorF2 for low-mass systems involving NSs and restrict themselves to only relatively small spins (dimensionless spin parameters $|\chi_{1,2}|\leq0.05$;~\citep{cbcCompanion}). 
We performed a grid-based calculation to determine the possible improvements in detectability provided by spins up to $|\chi_{1,2}|\leq0.1$.  
We find that including spins only marginally increases \rhorec/\rhonet~ (e.g., from 0.30 to 0.34 for $A=10^{-6}$).
The slight improvement is likely due to spins compensating somewhat at high frequencies for the biases in chirp mass (see \S~\ref{s:biases} and Appendix~\ref{s:derivation of phase shift}) induced at low frequencies by the NL effects.
Although we did not fully explore the effect of spins, our analysis suggests that measurements of the spin may be biased, which could have implications for population synthesis inferences~\citep{AstroImplications}.

Full spinning waveforms may increase the match further, but it is unlikely that they will recover a significant fraction of the lost \rhonet.
We conclude that spin may be important for studies of populations of marginally detectable sources with marginally relevant values of $A$. 
However, when $A$ is large, we see a dramatic reduction in our ability to recover signals even when using spinning PP waveforms.


\subsection{Biases}\label{s:biases}

\begin{figure}
    \includegraphics[width=1.0\columnwidth]{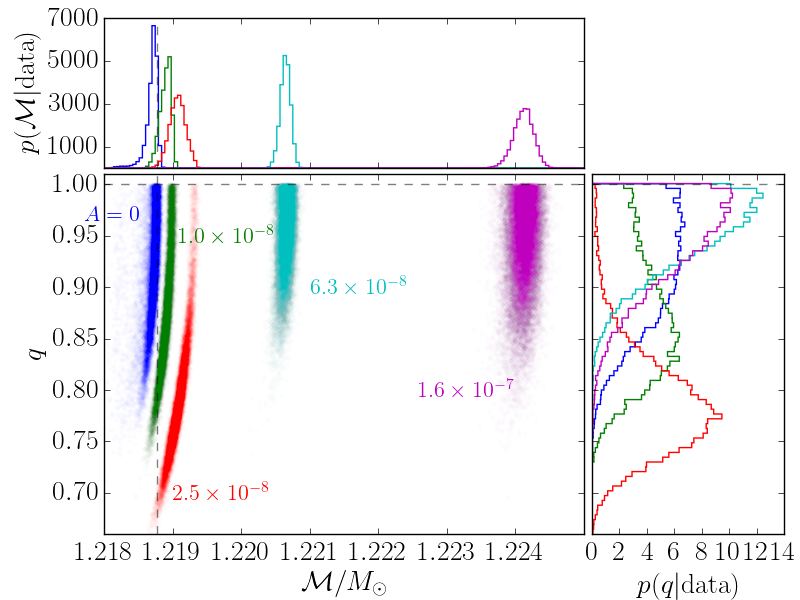}
    \includegraphics[width=1.0\columnwidth]{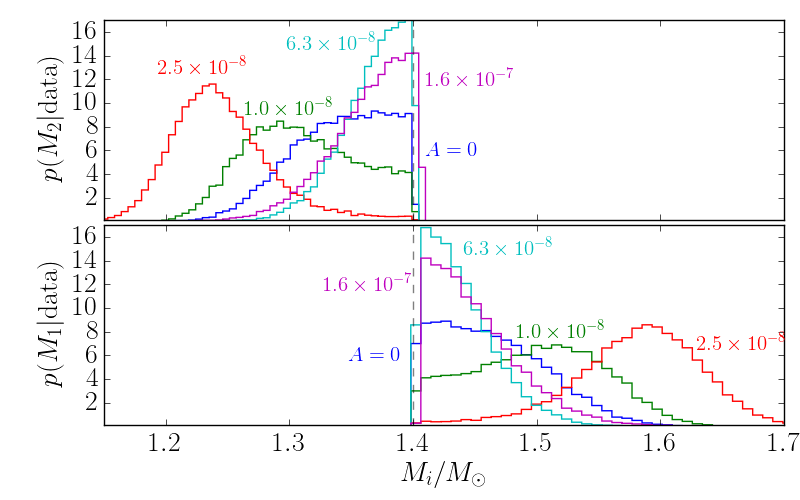}
    \caption{
        (top) Joint and marginal posterior distributions of $\mathcal{M}$ and $q$ for injected signals that include nonlinear tidal effects but are recovered using PP waveforms.  
        (bottom) Marginal distributions for the individual component masses, which are restricted to $M_1 \geq M_2$.
        The different curves show results for different values of $A$.
        We take $f_0=50\,\mathrm{Hz}$ and $n=0$, and inject the signals at $\rhonet\simeq25$.
        }
    \label{f:component masses}
\end{figure}

When $A$ is small, PP models fit the true waveform well and the posterior distributions are centered on the true values.
At $A\sim10^{-8}$ we begin to observe biases in the recovered parameters even though \Oppn~has decreased by only a few percent.  
This is sometimes called a ``stealth bias''~\citep{Cornish2011,Vallisneri2013}.
Figure~\ref{f:component masses} shows the joint and marginal posterior distributions of the chirp mass $\mathcal{M}$  and the mass ratio ($q=M_2/M_1$) as a function of $A$ for $n=0$, $f_0=50\textrm{ Hz}$ with $\rhonet\simeq25$.
Here and throughout the rest of this study, we follow the standard convention $M_1\geq M_2$ so that $0<q\leq1$, reflecting a symmetry under the interchange $M_1\leftrightarrow M_2$.
$\mathcal{M}$ is measured particularly well because it dominates the frequency evolution of inspirals~\citep{Peters1963}.
We observe a clear bias in $\mathcal{M}$ as $A$ increases.
This is because larger $A$ imply faster orbital decay, which can be confused with heavier systems that also decay faster.
Even at $A=10^{-8}$, we observe a statistically significant bias in $\mathcal{M}$ even though \Oppn~is essentially identical to the $A=0$ result.
Therefore, nonlinear tidal effects can bias parameter estimation even before they impact detection.
However, we note that although the bias  in $\mathcal{M}$ can be much larger than the statistical uncertainty, in absolute terms it remains small ($\lesssim1\%$) even for large values of $A$.

Nonlinear tides also introduce biases in the mass ratio  $q$, particularly when the impact on detectability is marginal. 
For $A\lesssim5\times10^{-8}$, $q$ is biased toward more asymmetric component masses.
This is because asymmetric systems also decay faster.
In fact, for large $f_0$, $q$ is biased so much that $\mathcal{M}$ is inferred to be \emph{smaller} than it really is (see Appendix~\ref{s:data}).
For our 1.4$M_\odot$-1.4$M_\odot$ system, we find that at $A\sim \textrm{ few}\times10^{-8}$  the larger mass may be inferred to be as much as 1.6$M_\odot$ and the smaller mass as little as 1.2$M_\odot$.
For different values of $n$, the bias in $q$ can be even more extreme than this.
Although we are not likely to misclassify a NS-NS binary as a NS-BH system for canonical 1.4$M_\odot$-1.4$M_\odot$ systems, there might be some confusion for masses near the maximal NS mass.

As Fig.~\ref{f:component masses} shows, the bias in $q$ is large for intermediate values of $A\sim 10^{-8}$ but small for $A\ll 10^{-8}$ and $A\gg 10^{-8}$.  
By contrast, we find that the bias in $\mathcal{M}$ increases nearly monotonically with $A$. 
Apparently, for $A\lesssim10^{-8}$, which corresponds to $\Delta\phi\lesssim1$ radian, the PP model can still approximate the data reasonably well, but only with a substantially biased $q$. 
We find that this trend holds for all values of $\f_0$ and $n$.
However, for $A\gg 10^{-8}$, \Oppn~decreases significantly and even though no set of PP parameters captures the data well, the true parameter values again offer the best fit (with the exception of $\mathcal{M}$, which remains biased at large $A$).

Despite the potential for biases, the posteriors for the component masses $M_1$ and $M_2$ almost always have some support near the true value, even if it corresponds to a long tail relative to the mode of the  distribution.
We also find that heavier systems with larger $\mathcal{M}$ (including NS-BH systems) are less biased by NL effects because $\Delta \phi \propto \mathcal{M}^{-10/3}$ [see Eqs. (\ref{e:phase shift}) and (\ref{e:Fm}) and the discussion in \S~\ref{s:parameterized model}].
Such systems have smaller $\Delta \phi$ because they decay faster and spend less time in the slow inspiral phase where nonlinear tides make their greatest impact.  Therefore, for the same $A$, the posteriors and odds ratios of NS-BH systems more closely resemble the PP model.

\begin{figure}
    \includegraphics[width=1.0\columnwidth]{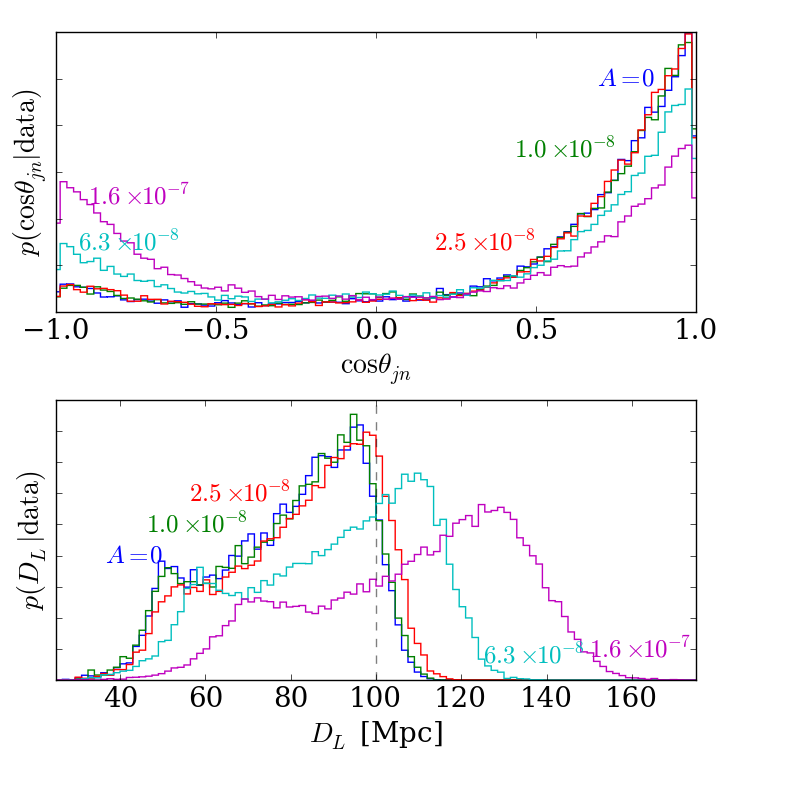}
    \caption{
        Posterior distributions for luminosity distance $D_L$ and inclination $\theta_{jn}$ for injected signals that include nonlinear tidal effects but are recovered using PP waveforms. 
        We take $f_0=50\,\mathrm{Hz}$, and $n=0$, and inject the signals at $\rhonet\simeq25$, corresponding to $D_L\simeq100$ Mpc.
        }
    \label{f:extrinsic}
\end{figure}

Nonlinear tides can also bias the luminosity distance $D_L$.  
In Fig.~\ref{f:extrinsic} we show the posterior distributions of $D_L$ and orbital inclination $\theta_{jn}$ (the angle between the system's total angular momentum and the line of sight to the source).
As we showed above, the PP model compensates for increasing $A$ by increasing $\mathcal{M}$.
However, systems with larger $\mathcal{M}$ are intrinsically more luminous and therefore are inferred to come from larger $D_L$.
Despite the bias, we find that the posterior distribution of $D_L$ is broad enough to cover the true value for our injections.

The other extrinsic parameters, such as $\theta_{jn}$ and source position, are unbiased by nonlinear tides.
This is because the phase shift affects both polarizations equally and these other extrinsic parameters depend primarily on the ratio of the two polarizations. 
Although not biased, the decrease in \Oppn~with increasing $A$ does broaden the posteriors of all  extrinsic parameters.

\section{measurability and model selection with nonlinear tides}\label{s:nlnl}

Having quantified the impact of neglecting nonlinear tidal effects in \S~\ref{s:nlgr}, we now consider how well they can be measured when they are included in the analysis. 
In \S~\ref{s:model selection} we evaluate the statistical evidence for their existence, and in \S~\ref{s:measurability} we assess how well  we can constrain the nonlinear tide parameters from the data.
To do this, we repeat the simulations in \S~\ref{s:nlgr} but now use a model that \emph{does} include the nonlinear effects when recovering the signal. 
We thereby obtain posterior distributions for $A$, $n$, and $\f_0$ as well as odds ratios \Onln comparing the nonlinear tide model to Gaussian noise.


\subsection{Model selection}\label{s:model selection}

By computing both \Oppn~and \Onln, we obtain an odds ratio comparing the two signal models $\ln\Onlpp = \ln\Onln - \ln\Oppn$.  
This provides a statistical measure of the evidence for each model.  
If \Onlpp~is large, the nonlinear (NL) model is favored.

\begin{figure}
    \includegraphics[width=1.0\columnwidth]{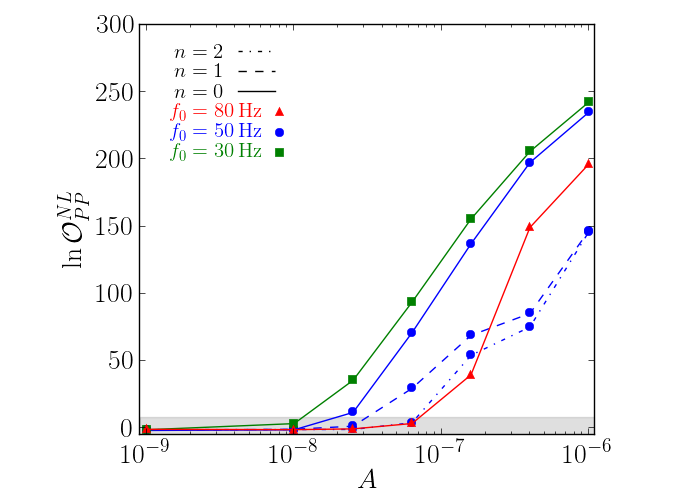}
    \caption{
        Odds ratio \Onlpp~for the same parameters as Fig.~\ref{f:nlgr logBSN}.
        }
    \label{f:nlnl logBNLGR}
\end{figure}

\begin{figure}
    \includegraphics[width=1.0\columnwidth]{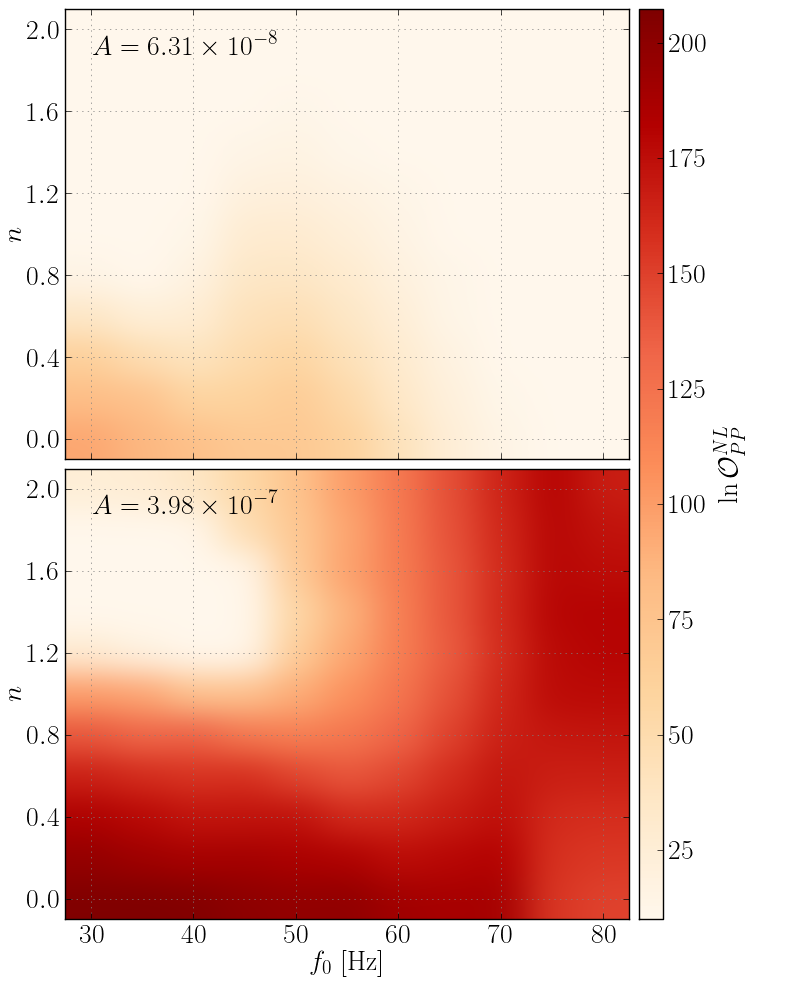}
    \caption{
        Surface plots of \Onlpp~as a function of $n$ and $f_0$ at (top panel) $A=6.3\times10^{-8}$ and (bottom panel) $A=4.0\times10^{-7}$. The signals are injected at $\rhonet\simeq25$.
        }
    \label{f:nlnl logBNLGR surf}
\end{figure}

In Fig.~\ref{f:nlnl logBNLGR} we show \Onlpp~as a function of $A$.
For $A\lesssim 10^{-8}$, we find $\Onlpp<1$ which implies that the model neglecting nonlinear tides is favored.
This is due to Occam's razor, which penalizes the more complicated models that include nonlinear tides because they do not match the data significantly better than the simpler models that ignore them.
Typically, the Occam factor corresponds to $\ln\Onlpp\sim-0.1$ and is not strongly dependent on \rhonet. 
For $\rhonet\simeq25$, this corresponds to less than 0.05\% of \Oppn.
However, when $A\gtrsim10^{-8}$, the NL models are strongly favored.
Comparing with Fig.~\ref{f:nlgr logBSN}, we see that $A\sim10^{-8}$ is also where \Oppn~begins to decrease.
This is not a coincidence. 
The NL models are able to reconstruct the signal equally well regardless of $A$ and thus $\Onln\simeq \textrm{constant}$.
Therefore, $\ln\Onlpp \simeq \textrm{ constant} - \ln\Oppn$ and the critical values of $A$ for detectability and model selection are the same.
Figure~\ref{f:nlnl logBNLGR surf} shows that the trend continues as a function of $n$ and $f_0$ as well.
Figures~\ref{f:nlnl logBNLGR surf} and \ref{f:nlgr logBSN surf} are inverses; areas that were ``hot'' become ``cold'' and vice versa.
Therefore, the regions of parameter space where the PP models fail correspond to the regions where the models with nonlinear tides are most favored.
It also means that we can recover all of the \rhonet~that is lost when nonlinear tides are neglected by using a more complete model.

\subsubsection{Tests of GR (with only linear tides)}\label{s:TIGER}

While it is clear that we can distinguish NL models from PP models for large $A$, it is also interesting to consider whether we can detect deviations from the PP model without the correct alternative model. 
Test Infrastructure for GEneral Relativity (\textsc{tiger})~\citep{Agathos2014,Li2012,Li2012b} is designed to answer exactly this question and computes odds ratios between the PP model and generic deviations from vacuum GR (\Otiger).
It does so by allowing the post-Newtonian (PN) coefficients to vary away from their GR predictions and computing the evidence for the modified models.
Furthermore, \textsc{tiger} is agnostic about the effects of linear tides and only considers $\f\lesssim400\Hz$~\citep{Agathos2015}.
In this way, it focuses on the early inspiral alone, during which the PP model is expected to be correct.
We used \textsc{tiger} to analyze a single injection ($A=1.6\times10^{-7}$, $f_0=50\,\Hz$, $n=2$) and observed large evidence for models allowing the first four PN coefficients to vary.
They correspond to $\ln\Otiger\simeq45$ when $\rhonet\simeq25$ and there is strong evidence in favor of the alternative hypothesis.
By comparison, when we use the NL model rather than \textsc{tiger} to recover the same injection we find $\ln\Onlpp\simeq53$.
We also note that $n=2$ corresponds to some of the smaller \Onlpp~observed; other parameters are likely to produce even larger evidence in favor of \textsc{tiger}'s alternative hypothesis.

Various studies have shown \textsc{tiger} to be insensitive to most uncertainties associated with compact binary coalescences and interferometric observatories (e.g., linear tides and calibration uncertainties; \citep{Agathos2013,Agathos2014}).
However, we find that nonlinear tide effects, if large and ignored, can fool the \textsc{tiger} machinery and suggest that GR is not the correct theory of gravity when, in fact, we have simply neglected relevant physics within the NSs.
To our knowledge, this is the first example of an effect that, if ignored, could fool \textsc{tiger}.
This therefore emphasizes the implicit assumption within the \textsc{tiger} analysis that all relevant physics has already been included in the model. 

In summary, although we only analyzed a single event with \textsc{tiger}, the results suggest that even imperfect models of the nonlinear tidal effects can significantly improve our ability to recover signals.

\subsection{Measurability}\label{s:measurability}

We found that neglecting nonlinear tides when $A\gtrsim10^{-8}$ can significantly hamper detection and bias parameter estimation.
Conversely, we found that if $A\gtrsim10^{-8}$, there will be strong statistical evidence for nonlinear tides.
We now consider how well we can measure the nonlinear parameters with data from a single event.

\begin{figure}
    \includegraphics[width=1.0\columnwidth]{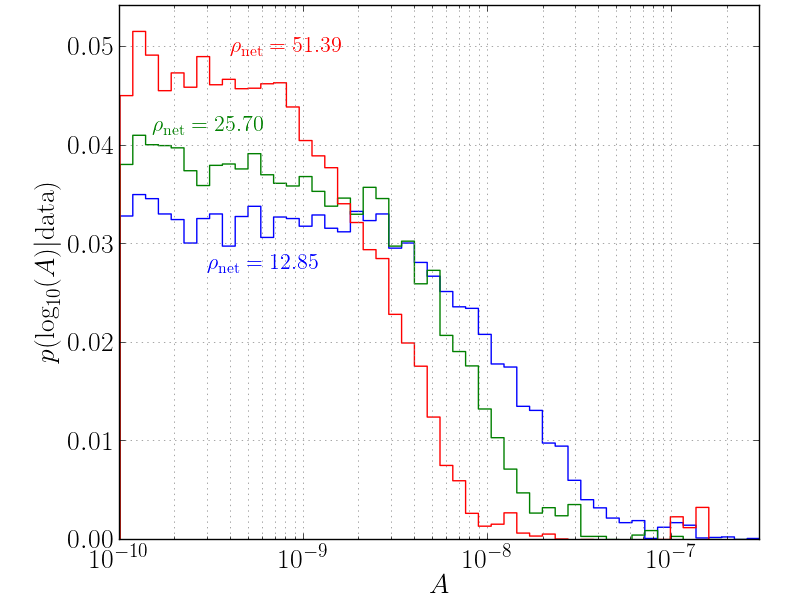}
    \caption{
        Posterior distributions for $A$ when the injected signal does not include nonlinear tide effects.
        }
    \label{f:grnl p(A|data)}
\end{figure}

We first evaluate what upper bound on $A$ is achieved when nonlinear effects are extremely small (i.e., for injected signals with $A\rightarrow 0$).
In Fig.~\ref{f:grnl p(A|data)} we show the posterior distributions of $A$ for different values of \rhonet~assuming a uniform prior for $\log A$. 
We find that the upper bound is near $A\sim10^{-8}$, with a slight decrease with increasing \rhonet.  This is not surprising given that at this $A$ the tidal effects begin to be noticeable (\S~\ref{s:nlgr}).  

\begin{figure}
    \includegraphics[width=1.0\columnwidth]{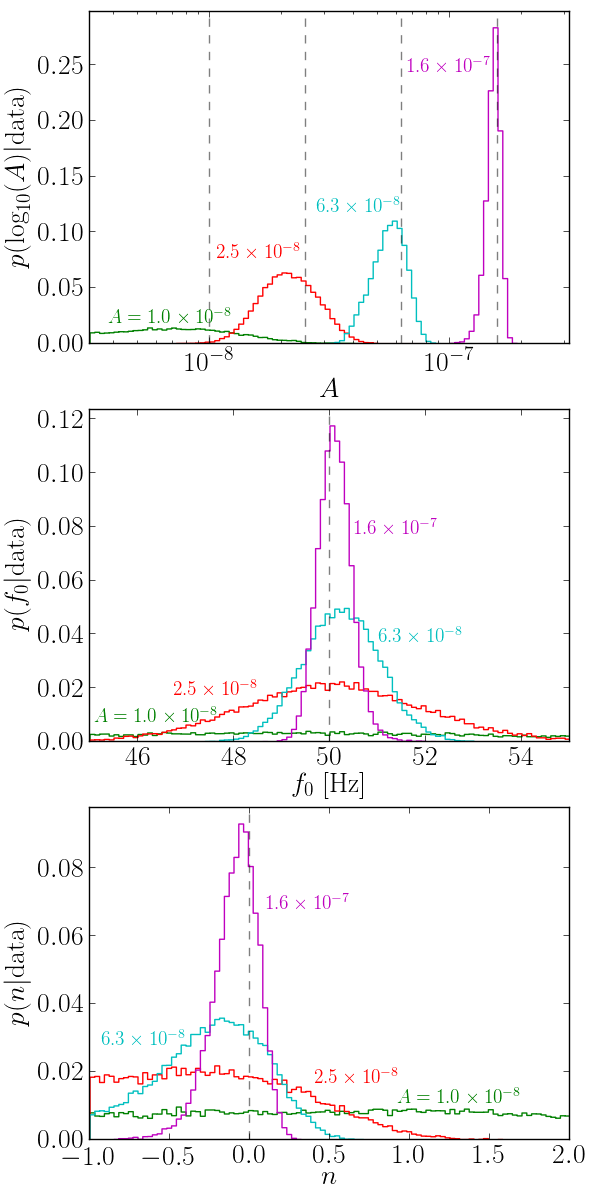}
    \caption{
        Measurability of (top) $A$, (middle) $f_0$, and (bottom) $n$ as a function of $A$.
        Vertical dashed lines show injected values.
        We take $f_0=50\,\mathrm{Hz}$ and $n=0$, and inject the signals at $\rhonet\simeq25$.
        }
    \label{f:nlnl p(A,n,f0|data)}
\end{figure}

In Fig.~\ref{f:nlnl p(A,n,f0|data)} we show the marginal posterior distributions for $A$, $n$, and $f_0$ for injections at $\rhonet\simeq25$. 
When $A\lesssim10^{-8}$, we cannot measure $n$ or $f_0$.
However, for $A\gtrsim10^{-8}$, we can measure both $n$ and $f_0$ to relatively high precision even at $\rhonet\sim12$.
Typically, we measure $A$ and $\f_{0}$ comparably, based on a comparison of the Kullback-Leibler divergence~\cite{kullback1951,O'Shaughnessy2013} from the prior to the posterior and the entropy of the posteriors.
Measuring $n$, however, requires either larger $A$ or \rhonet.

\begin{figure}
    \includegraphics[width=1.0\columnwidth]{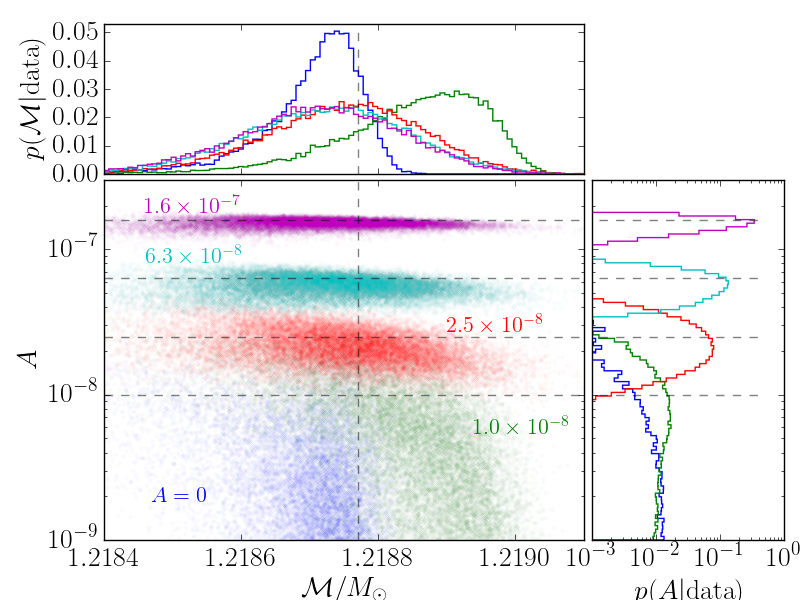}
    \caption{
        Joint and marginal posterior distributions of $\mathcal{M}$ and $A$ for various values of $A$.
        We take $f_0=50\,\mathrm{Hz}$ and $n=0$, and inject the signals at $\rhonet\simeq25$.
        }
    \label{f:correlations}
\end{figure}

There are also degeneracies among many of the parameters in our model. 
The strongest degeneracy is between $A$ and $\mathcal{M}$, which we show in Fig.~\ref{f:correlations}. 
When $A\sim10^{-8}$ and nonlinear tides are marginally detectable we find a negative correlation between $\mathcal{M}$ and $A$ (larger $\mathcal{M}$ favor smaller $A$ and vice versa). 
This is because a bias toward larger $\mathcal{M}$ shortens the  inspiral and thereby mimics the effects of the nonlinear tide.
When $A\gtrsim10^{-8}$,  the degeneracy between $\mathcal{M}$ and $A$ is present but truncated because $A\ll 10^{-8}$ is ruled out.

We also find degeneracies between the nonlinear tidal parameters.
$A$ has a strong positive correlation with $n$ because $\Delta\phi\propto A/(n-3)$. 
Interestingly, this forces $A$ to have a weak negative correlation with $\f_0$ because larger $A$ imply larger $n$, which then requires a smaller $\f_{0}$ to maintain roughly the same $\Delta\phi$.
However, this correlation breaks down for large $\f_{0}$,  because $\Delta \phi$ then depends more strongly on $\f_{0}$, weakening the correlation between $n$ and $A$ and strengthening the correlation between $\f_{0}$ and $A$.


\subsubsection{Dependence on component mass}\label{s:mass dependence}

So far, we have focused on only the leading order terms in our Taylor expansions of $A$, $\f_0$, and $n$ [see Eq. (\ref{e:taylor})].
However, our reconstructions also sampled the first order terms.
We do not find any strong correlations between the zeroth and first order terms.
Nonetheless, while most marginal posterior distributions for the first order terms are completely unconstrained, occasionally we observe weak constraints on $d\f_{0}/dm$ near the boundary of its prior.

If this result holds more generally and we are able to measure $\f_{0}$ as a function of component mass from a series of detections, we may be able to use $\f_{0}$ to make cosmological measurements using GWs alone.
This is because $\f_{0}$ provides an intrinsic frequency scale that gives a handle on the redshift of the otherwise conformal inspiral~\citep{Read2012,Mandel2014}.
Indeed, if we can measure $\f_{0}$ as a function of mass, we may extract both the redshift and the luminosity distance directly from the GW signal without recourse to an electromagnetic counterpart.
Similar approaches already exist in the literature including when one knows the NS equation of state~\cite{Read2012,DelPozzo15}, when the postmerger signal is observed~\citep{Messenger2014}, when the shape of the NS mass distribution is known~\cite{Taylor12a,Taylor12b}, and when no electromagnetic counterpart is found but there is a reliable galaxy catalog~\citep{DelPozzo12}.
Further studies will be needed, however, to test the usefulness of $\f_0$ and to evaluate the robustness of our saturation model.

We also carried out analyses in which we allow each body to have independent values of $A$, $\f_0$, and $n$ (as an alternative to the Taylor series expansions in component mass).  
Because there is a relatively weak dependence on mass ratio $q$ in the phase shift (Appendix~\ref{s:derivation of phase shift}) and because binary NS systems should have $q\sim1$, we find a strong degeneracy between $A_1$ and $A_2$.
Generally, the posterior supports large $A$ for one mass and small $A$ for the other, disfavoring nearly equal $A$ for both masses (even if the masses are similar). 
The Taylor expansion approach, by contrast, ensures similar values of $A$ for similar mass NSs. We therefore consider it a better method.
Most important, the weak constraints placed on the first order terms suggests that we capture most of the nonlinear tidal effects with just the zeroth order terms.


\section{Summary and Conclusions}\label{s:implications}

By constructing a parameterized model of the saturation of the $p$-$g$ instability in coalescing binary NSs, we explored how the instability might impact GW signals for current detector sensitivities. 
Our model contains three parameters ($A$, $f_0$, and $n$), where $A$ and $n$ determine the magnitude and frequency dependence of the nonlinear dissipation rate $\dot{E}_{\rm NL}$, and $f_0$ is the GW frequency at which the unstable modes saturate. 
Applying a full Bayesian analysis, we determined as a function of $A$, $\f_0$, and $n$ the extent to which nonlinear tidal effects: (1) influence the detectability of merger events, (2) bias binary parameters such as the chirp mass $\mathcal{M}$,  the mass ratio $q$, the component masses, and the luminosity distance $D_L$, and (3) can be measured.
We also examined, albeit in less detail, how the instability might be confused with NS spin and generic deviations from vacuum GR when a PP model is assumed at low frequencies.

We find that neglecting  nonlinear tidal effects can significantly impair our ability to detect events. 
For example, if $A\sim10^{-7}$, $n=0$, and $f=50$ Hz, we would lose $\simeq 30\%$ of \rhonet. 
This means that if we neglect nonlinear tides, we would miss $1-(0.70)^3\simeq 70\%$ of NS merger events. 
If $A\sim 10^{-6}$, $n=0$, and $f=50$ Hz,  and we neglect nonlinear tides, we would miss $\simeq 95\%$ of NS merger events. 
More generally, we find that nonlinear effects are detectable if $A\gtrsim 10^{-8}$.
An $A\sim 10^{-8}$ yields a phase shift relative to the PP waveform of $\Delta \phi \sim 1\textrm{ radian}$ and corresponds to, e.g., $N\sim 1$ ($\sim 100$) modes with $\omega_g/\omega_0\sim 10^{-3}$ ($\sim 10^{-4}$) saturating at $E_{\rm sat}\sim 0.1 E_{\rm break}$ [see Eq. (\ref{eq:Aparam})]. 
Although $N$ and $E_{\rm sat}$, and therefore $A$, are highly uncertain, values as large as $A\sim 10^{-6}$ and thus $\Delta \phi \sim 10^2\textrm{ rad}$ are a possibility (see  \S~\ref{s:physical mechanism} and \S~\ref{s:parameterized model}).
  
We also found that intrinsic parameter biases can be significant if nonlinear tidal effects are neglected.  
For example, we found that for $A\sim \textrm{ few}\times10^{-8}$, a $1.4M_\odot-1.4M_\odot$ NS-NS binary could be strongly biased to $1.6M_\odot-1.2M_\odot$.  
Interestingly, at this $A$ the loss in signal \rhonet~is relatively mild ($\lesssim 10\%$) and the PP waveform model would appear to be a good match to the data, an example of a ``stealth bias.''
For larger $A$,  the biases in many of the parameters tend to actually \emph{decrease} with increasing $A$ (the bias in $\mathcal{M}$ does not follow this pattern, however).  
Nonetheless, the quality of the PP model's match always worsens with increasing $A$.

We also used \textsc{tiger} to investigate whether we can detect deviations from the PP model without knowing the precise form of the nonlinear effects.
Although the evidence in favor of \textsc{tiger}'s alternative hypothesis is less than the evidence in favor of the exact nonlinear model, it does provide a significantly better match than the PP model if $A\gtrsim 10^{-8}$. 
This suggests that we may not need to know the precise form of the nonlinear effects in order to improve the match to the data.  
Moreover, it highlights the fact that neglected NS physics can produce apparent deviations from GR.

For heavier systems, such as NS-BH systems, nonlinear effects are significantly less important.  
This is because their orbits decay faster, giving the nonlinear tides less time to modify the inspiral.  
Therefore, for the same $A$, $\f_0$, and $n$, their waveform phase shifts are much smaller.   

Assuming that we observe a cosmological population of sources, nonlinear tides may provide a way to extract distance-redshift information directly from GW waveforms without identification of an electromagnetic counterpart.
This is because they provide a characteristic frequency $\f_0$ that breaks the otherwise conformal waveform. 
By measuring $\f_0$, we can extract the redshift directly and associate it with the corresponding $D_L$. 
Other studies of tidal effects have suggested similar approaches~\citep{Read2012,Mandel2014}.  
However, we will need to tightly constrain the possible values of $\f_0$ a priori in order to make such cosmological measurements.

Our study only analyzed single events and in the future it might be interesting to consider the impact of the $p$-$g$ instability on a population of sources. 
Such a study would benefit greatly from first improving the theoretical constraints on $A$, $n$, and $\f_0$.
A first-principles calculation of the saturation should therefore be very valuable.  
In addition to helping further assess the potential impact of nonlinear tides, it might also aid parameter estimation and detection pipelines by reducing the amount of parameter space that must be searched. 
Although a full saturation calculation would be ideal, even relatively small improvements could be useful, such as confirming the expected growth rates of the $p$-$g$ instability and more accurately determining the instability threshold and number of unstable modes.


\section{acknowledgments}\label{s:acknowledgements}

The authors would like to thank Scott Hughes for useful discussions and encouragement throughout this project as well as Chris Van Der Broek, Michalis Agathos, Richard O'Shaughnessy, and the referee for their comments and suggestions on the draft.
R.E. and N.W. were supported in part by NASA ATP Grant No. NNX14AB40G. 
The authors also acknowledge the support of the National Science Foundation and the LIGO Laboratory.
LIGO was constructed by the California Institute of Technology and Massachusetts Institute of Technology with funding from the National Science Foundation and operates under cooperative agreement PHY-0757058. 
The authors would like to acknowledge the LIGO Data Grid clusters. 
Specifically, we thank the Albert Einstein Institute in Hannover, supported by the Max-PlanckGesellschaft, for use of the Atlas high-performance computing cluster.


\bibliography{arxiv_refs}


\appendix

\section{Phase shift due to the nonlinear tide}\label{s:derivation of phase shift}

We compute the tidal phase shift $\Delta \phi(f)$ relative to the nonspinning PP model using a zeroth order PN expansion. 
We expect that higher order PN terms will simply add to the PP result without significantly modifying the effects from nonlinear tidal interactions.
Moreover, any correction from higher order PN terms will be small compared to the zeroth order term since the phase shift accumulates predominantly at low frequencies ($f\lesssim100\,\mathrm{Hz}$).

We assume a circular, quasi-Keplerian orbit that loses energy due to gravitational radiation and dissipative tidal interactions (between star 1 and star 2)
\begin{equation}
\label{e:Ebalance}
   \dot{E}_\mathrm{orb} = - \dot{E}_{\rm gw} - \dot{E}_1 - \dot{E}_2, 
\end{equation}
where
\begin{equation}
  \dot{E}_\mathrm{orb} = - \frac{G^{2/3}\pi^{2/3}\mathcal{M}^{5/3}\dot{\f}}{3\f^{1/3}},
\end{equation}
$\mathcal{M}=(M_1 M_2)^{3/5}/(M_1+M_2)^{1/5}$ is the chirp mass, $\f=\Omega/\pi$ is the GW frequency, $\Omega=[G(M_1+M_2)/a^3]^{1/2}$ is the Keplerian frequency, and~\citep{Peters1963}
\begin{equation}
  \dot{E}_{\rm gw} 
                =  \frac{32 \pi^{10/3}}{5} \frac{G^{7/3}\mathcal{M}^{10/3}}{c^5} \f^{10/3}. 
\end{equation}

We model the dissipation due to the tide raised in $M_1$ by $M_2$ as
\begin{equation}
  \dot{E}_1 =  \Gamma_1 N_1 E_\mathrm{sat,1}
\end{equation}
(and similarly for the tide raised in $M_2$ if both objects are NSs), where $\Gamma$ is the growth rate of the instability, $N$ is the number of unstable modes, and $E_\mathrm{sat}$ is the energy at which the unstable modes saturate. As we describe in \S~\ref{s:physical mechanism},
\begin{equation}
  \Gamma_1 = 2 \lambda_1 \epsilon_1 \omega_{0,1} = 2 \lambda_1 \frac{M_2}{M_1} \left(\frac{R_1}{a}\right)^3 \omega_{0,1},
\end{equation}
\begin{equation}
  E_{\mathrm{sat},1} = \beta_1 E_{\mathrm{break},1} = \beta_1 \left( \frac{\omega_{g,1}}{\Lambda_{g,1} \omega_{0,1}} \right)^2 E_{0,1},
\end{equation}
where $\omega_0^2 = GM/R^3$ and $E_0=GM^2/R$. Thus,
\begin{multline}\label{e:edot1}
    \dot{E}_1 = 2 \pi^2 \frac{M_1 M_2}{M_1+M_2} (GM_1)^{2/3} \\
        \times \left[ \omega_{0,1}^{-1/3} \left(\frac{\omega_{g,1}}{\Lambda_{g,1} \omega_{0,1}}\right)^2 \beta_1 N_1 \lambda_1 \right] \f^2. 
\end{multline}
As the orbit decays, the fraction of the breaking amplitude at which the instability saturates ($\beta$) may increase and there may be more unstable modes ($N$).
Therefore, we expect these parameters to vary with  frequency and for simplicity we assume
\begin{equation}
\beta_1 N_1 \lambda_1 = \left[\beta_1 N_1 \lambda_1\right]_{\rm ref} \left(\frac{\f}{\fref}\right)^{n_1} \Theta_1,
\end{equation}
i.e.,  a power law dependence with a sudden onset of the dissipation at $f=f_{0,1}$ as captured by the Heaviside function $\Theta_1 = \Theta\left(f-f_{0,1}\right)$ (the latter assumption is motivated by the rapid growth rates relative to the inspiral rate as described in \S~5.4 of W16).  We define the magnitude of $\beta_1 N_1 \lambda_1$ relative to the value at an arbitrary reference frequency $\fref$. Throughout our study we set $\fref=100 \textrm{ Hz}$ (for both star 1 and star 2).  Then
\begin{equation}
   \dot{E}_1 =  \frac{\left(2GM_1\right)^{2/3}M_1 M_2}{M_1+M_2}   \left(\pi\fref\right)^{5/3} 
    A_1 \left(\frac{\f}{\fref}\right)^{2+n_1} \Theta_1, 
\end{equation}
where 
\begin{eqnarray}
  A_1 &=& \left(\frac{2\pi \fref}{\omega_{0,1}}\right)^{1/3} \left(\frac{\omega_{g,1}}{\Lambda_{g,1} \omega_{0,1}}\right)^2  \left[\beta_1 N_1 \lambda_1\right]_{\rm ref}
  \nonumber \\
 &\simeq& 4\times10^{-9}  \left(\frac{\omega_{g,1}}{10^{-4}\Lambda_{g,1} \omega_{0,1}}\right)^2\left[\beta_1 N_1 \lambda_1\right]_{\rm ref}
 \label{eq:A1value}
\end{eqnarray}
is a dimensionless amplitude parameter that depends on the equation of state and how the instability saturates. The three parameters of our saturation model are therefore $A_1$, $n_1$, and $f_{0,1}$ for star 1 and similarly for star 2.  We expand each of these parameters about a $1.4 M_\odot$ reference mass as, e.g., $A_1 = A^{(0)} + A^{(1)}\left(M_1-1.4 M_\odot\right)+\cdots$, where the $A^{(i)}$ are the same for both NSs.  In practice, we keep only the zeroth and first order terms in our model.

Equation (\ref{e:Ebalance}) then implies
\begin{equation}
  \dot{\f} = 3\pi \fref^2 x^{7/3}\left[Bx^{4/3} 
            +  C_1 x^{n_1}+ C_2 x^{n_2}\right]
\end{equation}
where $x=\f/\fref$,
\begin{eqnarray}
B&=& \frac{32}{5}\left(\frac{G\mathcal{M}\pi \fref}{c^3}\right)^{5/3},\\
C_1&=&\left(\frac{2M_1}{M_1+M_2}\right)^{2/3} A_1 \Theta_1,
\end{eqnarray}
and similarly for $C_2$.  The phase of the GW signal $d\phi = 2\pi f dt = 2\pi \f d\f / \dot{\f}$ and
\begin{equation}
  \phi(\f) = 
  \frac{2}{3}\int\limits_0^{\f/\fref} 
\frac{x^{-4/3} dx}
{Bx^{4/3} +C_1 x^{n_1}+C_2x^{n_2}}.
\end{equation}
For typical NS parameters, $B\sim 10^{-4}$ and $B \gg C_{1,2} \approx A_{1,2}$ as long as
\begin{equation}
  \left[\beta N \lambda\right]_{\rm ref} \ll 10^5\left(\frac{10^{-4}\Lambda_{g} \omega_{0}}{\omega_{g}}\right)^{2}.
\end{equation}
which we expect to be satisfied.  Thus, the tidal decay due to gravitational radiation always strongly dominates and we can expand the $\phi(f)$ integrand as a power series. The phase shift relative to the PP waveform is therefore
\begin{eqnarray}
  \Delta \phi (f) &\simeq& -\frac{2}{3B^2}  \int\limits_0^{\f/\fref} dx \left[C_1 x^{n_1-4}+C_2x^{n_2-4} \right]
  \nonumber \\
  &\simeq& 0.4 \left(\frac{\mathcal{M}}{1.2M_\odot}\right)^{-10/3}\left(\frac{C_{1,2}}{10^{-8}}\right) \left[\frac{x_0^{n-3}   -x^{n-3}}{n-3}\right]
  \textrm{ rad},
  \nonumber \\
 \end{eqnarray}
where in the second line $x_0=f_0/\fref$, and we assumed $n<3$ and $M_1=M_2$.
The phase shift is negative which means that the orbit reaches a given frequency in fewer orbits than in the PP model. 


\section{Priors on the Model Parameters}\label{s:priors}

We use a Bayesian framework to compute the evidence and posterior distributions. 
In Table~\ref{t:priors}, we list the priors on all our model parameters. 
Only a few corner cases produced posteriors which railed against these priors, and those only manifested for extremely biased values of $q$.
In these few cases, the lower bound on $M_2$ acted as an effective bound on $q$.

\begin{table*}
  \caption{Prior distributions for the model parameters}
  \begin{center}
    \begin{tabular}{cc|c|c|l}
        \multicolumn{2}{c|}{Parameter} & Minimum & Maximum & Distribution \\
        \hline
        \hline
        \multirow{6}{*}{PP} & $M_1$                 & 1$M_\odot$ & 10$M_\odot$ & $dN \propto dM_1$             \\
                            & $M_2$                 & 1$M_\odot$ & 10$M_\odot$ & $dN \propto dM_2$             \\
                            & $D_L$                 & 0 Mpc      & 300 Mpc     & $dN \propto D_L^2 dD_L$       \\
                            & $\cos\theta_{jn}$     & -1         & 1           & $dN \propto d\cos\theta_{jn}$ \\
                            & $\alpha$              & 0          & 2$\pi$      & $dN \propto d\alpha$          \\
                            & $\cos\delta$          & -1         & 1           & $dN \propto d\cos\delta$      \\
        \hline
        \multirow{6}{*}{NL} & $A(1.4M_\odot)$                & $10^{-10}$                 & $10^{-5}$                 & $dN \propto d\log A$      \\
                            & $\frac{1}{A}dA/dm(1.4M_\odot)$ & -1$M_\odot^{-1}$           & 1$M_\odot^{-1}$           & $dN \propto d(\log A/dm)$ \\
                            & $f_0(1.4M_\odot)$              & 10$\,\mathrm{Hz}$          & 100$\,\mathrm{Hz}$        & $dN \propto df_0$         \\
                            & $df_0/dm(1.4M_\odot)$          & -10$\,\mathrm{Hz}/M_\odot$ & 10$\,\mathrm{Hz}/M_\odot$ & $dN \propto d(df_0/dm)$   \\
                            & $n(1.4M_\odot)$                & -1                         & 3                         & $dN \propto dn$           \\
                            & $dn/dm(1.4M_\odot)$            & -1$M_\odot^{-1}$           & 1$M_\odot^{-1}$           & $dN \propto d(dn/dm)$     \\ 
    \end{tabular}
  \end{center}
  \label{t:priors}
\end{table*}


\section{Correlations when $n=2$}\label{s:data}

In the main text we show the correlation between $\mathcal{M}$, $q$, and $A$ only for the $n=0$ case (see Figs. \ref{f:component masses} and \ref{f:correlations}).   However, as we show here, the trends are somewhat different when $n=2$. Thus, the correlations can change their behavior depending on the values of the injected parameters.

\begin{figure*}
    \includegraphics[width=1.0\columnwidth]{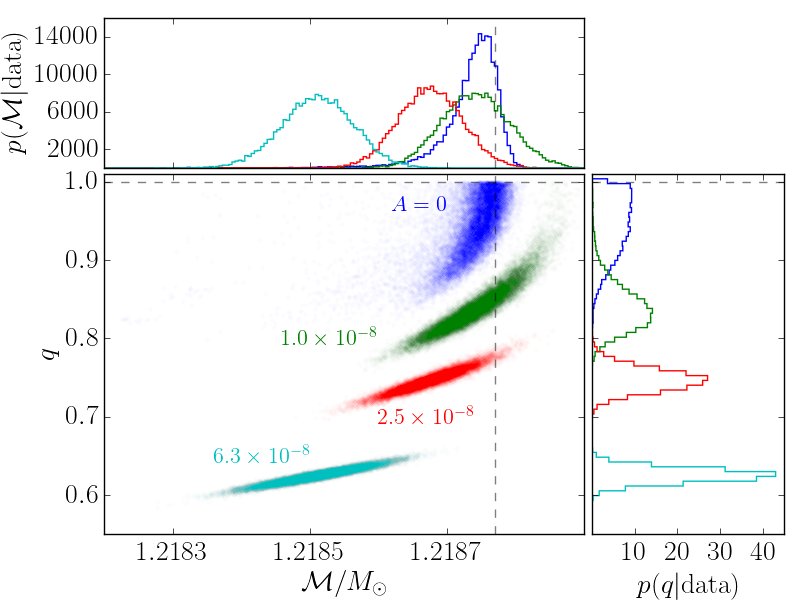}
    \includegraphics[width=1.0\columnwidth]{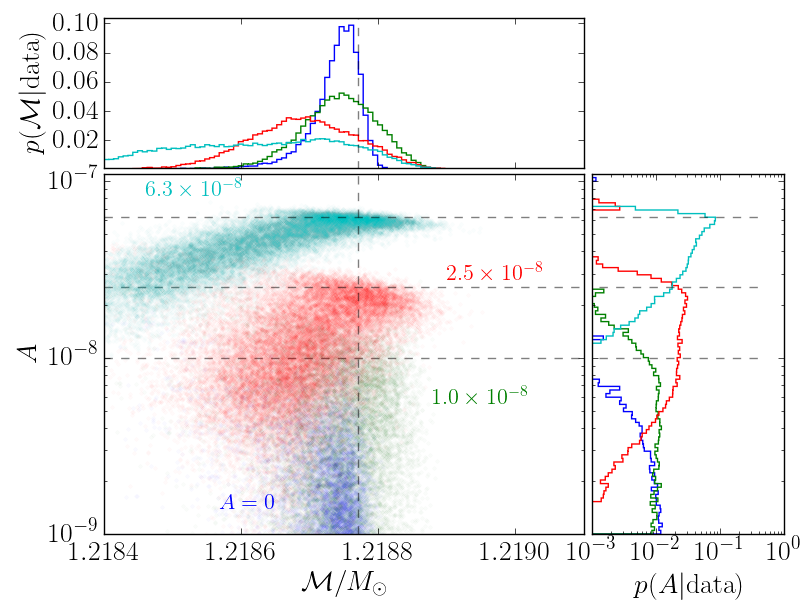}
    \caption{
    Joint and marginal posterior distributions of $\mathcal{M}$, $q$, and $A$ for various values of $A$.
        We take $f_0=50\,\mathrm{Hz}$ and $n=2$, and inject the signals at $\rhonet\simeq50$.
}
    \label{f:mchirp 2 50 50}
\end{figure*}

The left panel of Fig.~\ref{f:mchirp 2 50 50} shows the joint and marginal distributions for $\mathcal{M}$ and $q$ for injections with $n=2$, $\f_0=50\,\Hz$, and $\rhonet\simeq50$.
Unlike  in Fig.~\ref{f:component masses} where $\mathcal{M}$ is biased to larger values as $A$ increases,  here we see that $\mathcal{M}$ is biased to smaller values as $A$  increases.
This is because the bias in $q$ is much stronger and pushes the posterior backward along the degeneracy between $\mathcal{M}$ and $q$~\citep{Baird2013}. 

The right panel of Fig.~\ref{f:mchirp 2 50 50} shows the correlation between $\mathcal{M}$ and $A$ when $n=2$.  We see that it is reverse from the $n=0$ case shown in Figure~\ref{f:correlations}. In particular, larger $A$ imply larger $\mathcal{M}$. This is because at smaller $A$, the model compensates with a more asymmetric $q$ and a decrease in $\mathcal{M}$.

\end{document}